\documentstyle[11pt]{article}
\def\ba{\begin{eqnarray}}
\def\ea{\end{eqnarray}}

\def\lb{\label}
\def\be{\begin{equation}}
\def\ee{\end{equation}}

\begin{document}
\title{About apparent superluminal drives in generic gravity theories}
 \author{ Juliana Osorio Morales \thanks{Instituto de Matem\'atica Luis Santal\'o (IMAS), UBA CONICET, Buenos Aires, Argentina
juli.osorio@gmail.com.} and Osvaldo P. Santill\'an \thanks{Instituto de Matem\'atica Luis Santal\'o (IMAS), UBA CONICET, Buenos Aires, Argentina
firenzecita@hotmail.com and osantil@dm.uba.ar.}}

\date {}
\maketitle

\begin{abstract}
As is well known, there exists warp drives in GR, such as the Alcubierre bubbles, which achieve an apparent faster than light travel \cite{alcubierre}.  A result due to Gao and Wald \cite{gaowald} suggests that such a travel is unlikely
for GR with matter satisfying both the Null Energy  and the Null Generic Conditions. There exists a generalization of this statement
due to Galloway, that ensures that the Gao-Wald result is true regardless the underlying gravity model, unless there exists at least one inextendible
null geodesic with achronal image in the space time (a null line). The proof of this proposition is based on techniques of causal theories, and has never been released.
In the present work an independent proof of this result is presented by use of the Raychaudhuri equation, and avoiding several technical complications described along the text. Some consequences of these affirmations are discussed at last section, in particular their potential use in problems of causality.

\end{abstract}

\section{Introduction}
After the introduction of the \emph{Alcubierre bubble} \cite{alcubierre} or the \emph{Krasnikov tube} \cite{krasnikov}, there has been a growing interest in
the topic of time advance in General Relativity as well as in modified theories of gravity.  The Alcubierre bubble is a space time in which it is possible to make a round trip from two stars $A$ and $B$ separated by a proper distance $D$ in such a way that a fixed observer at 
the star $A$ measures the proper time for the trip as less than $2D/c$. In fact, the duration of this travel can be made arbitrary small. This fact does not indicate that the observers travel faster than light, as they are traveling
inside their light cone. The Alcubierre constructions employ the fact that, for two comoving observers in an expanding universe, the rate of change of the proper distance to the proper time may be larger than $c$ or much more smaller, if there is contraction instead of expansion. The Alcubierre space time is Minkowski almost everywhere, except at a bubble around the traveler which endures only for a finite time. This bubble is specially designed for making the proper time of the trip measured by an observer at the star $A$ as small as possible. Details can be found in \cite{alcubierre}.

The examples described above are of physical interest, but the precise  definition of time advance is indeed very subtle \cite{olum}.  A careful definition of time advance was introduced in \cite{olum}. In this reference, a space time which appears to allow time advance was constructed, but it was proven that it is in fact
the flat Minkowski metric in unusual coordinates. This suggests that to conclude time advance by simple inspection of the metric may be misleading. The definition of time advance for general space times is involved, and discussions about this can be found in the works \cite{otras}-\cite{otras4} and references therein. However, for  space times that are Minkowski outside a tube or a bubble such as Alcubierre or Krasnikov space times, the notion of time advance is easier to understand. The common point in all these constructions is the existence of a causal path going from two points $(t_1, x_1)$ to $(x_2, t_2)$ even though that $x_2-x_1>t_2-t_1$. The role of the tube or the bubble is to provide a local deformation of the space time in a region $K$, which is essential for this path to exist. The path in fact crosses that region $K$. By use of some results due to Tipler and Hawking \cite{tipler1}-\cite{tipler3}, it can be shown that all these examples violate the Null Energy Conditions at least in some part of this region. 

The results just described raise the question of whether time advance could hold in theories which do not violate the Null Energy Conditions.
In this context, a theorem due to Gao and Wald \cite{gaowald} may be relevant.  Its  statement is the following.
\\

\emph{Gao-Wald proposition:} Consider a null geodesically complete space time ($M$, $g_{\mu\nu}$) for which the Null Energy and Null Generic Conditions are satisfied. Then,  given a compact region $K$, there exists a compact $K'$ containing $K$
such that for any pairs of points $p, q\notin K'$ and  $q$ belonging to $J_+(p)-I_+(p)$, no causal curve $\gamma$ connecting both points  intersects $K$.
\\

The relevance of this theorem is as follows. Assume that one is intended to deform a given space time $M$ in a region $K$, similar perhaps to a bubble, in order to construct a path passing through $K$ and connecting two points that otherwise would be causally disconnected. The theorem states that this is not possible if the Null Energy Conditions and Null Generic Conditions are satisfied, unless the points are inside the larger region $K'$. This may constitute a sort of no go theorem. The problem is that there is no control over the size of the region $K'$. If the region $K'$ results infinitely large, this theorem loses its power. For this reason, this result should be considered only as a weak version  of a no go result.

Recall that the Null Energy Condition is fulfilled if and only if $T_{\mu\nu}$ satisfies $T_{\mu\nu}k^\mu k^\nu\geq 0$ for every null vector $k^\mu$ tangent to any null geodesic $\gamma$. This implies, in the context of General Relativity, that $R_{\mu\nu}k^\mu k^\nu\geq 0$ \cite{Wald}-\cite{penrose}. On the other hand, the Null Generic Condition means that $k_{[\alpha} R_{\beta]\sigma\delta[\epsilon}k_{\gamma]}k^\sigma k^\delta\neq 0$ for some point in the geodesic $\gamma$. Both conditions automatically imply that any null geodesic $\gamma(\lambda)$ possesses at least a pair of conjugate points $p$ and $q$, if it is past and future inextendible,  see \cite[Proposition 9.3.7]{Wald}.

 The results described above hold in the context of General Relativity, and should not be extrapolated to modified gravity theories without further analysis. However, at the footnote 1 of reference \cite{gaowald} it is commented that there exist a proposition due to Galloway that can be expressed as follows.
\\

\emph{Galloway's proposition:} Consider a space time $M$ in which every inextensible null geodesic can be deformed to a time like curve (this means all these geodesics contain at least two conjugate points, see proposition 1 below). Then,  given a compact region $K$, there exists a compact $K'$ containing $K$
such that for any pairs of points $p, q\notin K'$ and  $q$ belonging to $J_+(p)-I_+(p)$, no causal curve $\gamma$ connecting both points  intersects $K$.
\\

Note that this proposition does not employ any particular gravity model, neither impose conditions about the matter content or geodesic completeness. It should be emphasized that the presence of an inextendible null geodesic without conjugate points does not insure that the no go Gao-Wald theorem is avoided, it is a necessary but not sufficient condition. Nevertheless, it is clear  that the  only hope to avoid the Gao-Wald result is to find a scenario containing at least one of such curves. These null curves exist for instance, in the Alcubierre space time \cite{alcubierre}. This follows from the fact that, outside the bubble, the space is Minkowski and all the null geodesics in this region which do not cross the bubble have all obviously achronal images. Thus, the Alcubirre space time does not contradict these two propositions.

As far as the authors know, the Galloway theorem is a result of causal techniques applied to gravity models, and its proof has never been released. The motivation of the present letter is to obtain a similar result, but with a proof based in the properties of the Raychaudhuri equation. This will constitute an independent proof of the Galloway result. The reward is that some continuity properties of conjugate points along congruence of geodesics are proven, which are not evident from the Galloway statement.

The present work is written is self-contained manner. The advantage of this is that the text becomes more readable. The disadvantage is that the original contribution and the known results may be mixed. This distinction will be emphasized along the text to avoid credit confusion. The organization of the present work is as follows. In section 2 some generalities about conjugate points in generic space times are discussed. In addition, certain topological issues related to the light cones in space times 
are also presented. The presentation is not exhaustive, but focused in the aspects more relevant for our purposes. In section 3, a continuity lemma of fundamental importance for proving the Gao-Wald theorem is presented. In section 4, it is shown that this continuity lemma still holds even if the hypothesis of the Gao-Wald theorem are erased. In particular, a result similar to the Galloway theorem is found. A discussion about the possible applications is presented at the end.

\section{Generalities about conjugate points and a continuity argument}
As discussed above, the Gao-Wald theorem relies on the notion of conjugate points. It may be important to recall some of their basic properties,  following the references \cite{Wald}-\cite{penrose}.
\subsection{Null geodesics and conjugate points}
In the following $M$ always denotes a paracompact space time. It will be assumed the existence of a globally defined time like future pointing vector  $t_\mu$ on it. 
Given a point $p$ in  ($M$, $g_{\mu\nu}$), a point $q$ in $J_+(p)-I_+(p)$ is said to be conjugated to $p$ if the following holds.
Consider a null geodesic $\gamma(\lambda)$ emanating from $p$, together with the associated differential equation
\be\lb{smile}
\frac{d^2A^\mu_\nu}{d\lambda^2}=-R^\mu_{\alpha\beta\gamma}k^\alpha k^\beta A^\gamma_\nu,
\ee
supplemented with the following initial conditions
$$
A^\mu_\nu|_p=0,\qquad \frac{dA^\mu_\nu}{d\lambda}\bigg|_p=\delta^\mu_\nu.
$$
Here $\lambda$ is the affine parameter describing $\gamma(\lambda)$ and $k^\mu$ is a vector tangent to the curve $\gamma(\lambda)$,  and satisfying the following conditions
\be\lb{norma}
 k^\mu k_\mu=0, \qquad k^\mu t_\mu=-1.
\ee
The point $q=\gamma(\lambda_0)$ is said to be conjugated to $p$ if and only if $$\det(A_\mu^\nu(\lambda_0))=0.$$ 
 The matrix $A^\mu_\nu(\lambda)$ has the following interpretation: the components $A^{\mu}_{\nu}$ are the coefficients of the Jacobi field $\eta^{\mu}$ along $\gamma$, i.e, 
$$
\eta^\mu(\lambda)=A^\mu_\nu(\lambda) \frac{d\eta^\nu}{d\lambda}\bigg|_0,\qquad  \eta(0)|_p=0.
$$
The equation (\ref{smile}) implies that $\eta(\lambda)$ satisfies the Jacobi equation (hence the name) on $\gamma$ given 
by
\be\lb{jacobi}
\frac{d^2\eta^\mu}{d\lambda^2}=-R^\mu_{\alpha\beta\gamma}k^\alpha k^\beta \eta^\gamma.
\ee
The classical definition of a conjugate point $q$ to $p$ is the existence of a solution $\eta^\mu(\lambda)$ of the Jacobi equation such that $\eta^\mu(0)=0$ and $\eta^\mu(q)=0$. 
Clearly, the fact that $\det(A^\mu_\nu(\lambda_0))=0$ implies that there exist some initial conditions such that $\eta^\mu(q)=0$, thus $q$ is a conjugate point to $p$ in the usual sense.  For further details see \cite[Section 9.3]{Wald}.

There is no guarantee that there exists a point $q$ conjugate to a generic point $p$ for a given space time ($M$, $g_{\mu\nu}$). In addition, there might exist two or more different points $q$ and $s$ conjugate to $p$, joined to $p$ by different geodesics.

The study of conjugate points has been proven to have many applications in Riemannian and Minkowski geometry. It is well known that, in Riemannian geometry, a geodesic $\gamma(\lambda)$ starting at a point $p=\gamma(0)$ and ending at a point $r=\gamma(\lambda_0)$ is not necessarily length minimizing if there is a conjugate point $q=\gamma(\lambda_1)$ to $p$ such that $\lambda_1<\lambda_0$. The presence of a conjugate point in the middle usually spoil the minimizing property. 
For time like geodesics in Minkowski geometries, the proper time elapsed to travel between $p$ and $r$ is not necessarily maximal if there is a conjugate point in the middle. For null geodesics, there is an important result which will be used below, see \cite[Theorem 9.3.8]{Wald}.
\\

\emph{Proposition 1:} Let $\gamma$ a smooth causal curve and let $p, r\in \gamma$. Then there does not exist a smooth one parameter family of causal curves $\gamma_s$ connecting both points, such that $\gamma_0=\gamma$ and such 
that $\gamma_s$ are time like for $s>0$ if and only if there is no conjugate point $q$ to $p$ in $\gamma$.
\\

By reading this statement as a positive affirmation, it is found that if a null curve connecting $p$ and $r$ can be deformed to a time like curve, then there is a pair of conjugate points in between and, conversely, if there is  such pair, the curve can be deformed to a time like one. An inextendible causal curve that has no conjugate points has  achronal image. These are called null lines in the literature \cite{null}.
 
The matrix $A_\mu^\nu(\lambda)$ defined by equation (\ref{smile}) takes values which depend on the choice of the null geodesic $\gamma$. For this reason it may be convenient to denote it as $(A_\gamma)^\mu_\nu$. 
The same follows for the quantity
\be\lb{marolf}
G_\gamma(\lambda)=\sqrt{\det A_\gamma(\lambda)},
\ee
which also vanish at both $p$ and $q$. Note that the initial conditions below (\ref{smile}) imply that $A_\mu^\nu\sim \lambda \delta_\mu^\nu$ is positive for points in $J_+(p)-I_+(p)$  close enough to $p$, and so it is  $\det A_\gamma>0$. If there is a change of sign, then a conjugated point $q$ has been reached. Thus, for studying the first conjugate point $q$ to $p$, the square root in the definition of $G_\gamma(\lambda)$ does not pose a problem.
The equation (\ref{smile}) implies that $G_\gamma(\lambda)$ satisfies the following second order equation \cite{gaowald},\cite{marolf}
\be\lb{smile2}
\frac{d^2G_{\gamma}}{d\lambda^2}=-\frac{1}{2}[\sigma_{\mu\nu}\sigma^{\mu\nu}+R_{\mu\nu} k^\mu k^\nu]G_{\gamma},
\ee
and that $G_\gamma(0)=0$ and $G_\gamma(\lambda_0)=0$. These two values correspond to the points $p$ and $q$. Here
$\sigma_{\mu\nu}$ denotes the shear of the null geodesics emanating from $p$.
The last is an equation of the form
$$
\frac{d^2 G_\gamma}{d\lambda^2}=-p_\gamma(\lambda) G_\gamma.
$$
Near the point $p$ the initial conditions in (\ref{smile}) and the Jacobi formula for a determinant derivative imply that
\be\lb{inito}G_\gamma(0)=0, \qquad \frac{dG_\gamma(0)}{d\lambda}=0.\ee
Then, if $p_\gamma(\lambda)$  is $C^{\infty}$, by taking derivatives of equation (\ref{smile2}) with respect to $\lambda$ it may be shown that
$$
\frac{d^n G_\gamma(0)}{d\lambda^n}=0,
$$
for every value of $n$. This suggest that $G_\gamma(\lambda)$ may not analytical at the point $\lambda=0$. In other words, an attempt to solve equation (\ref{smile2}) with the initial conditions (\ref{inito}) may result in the trivial solution. However, $G_\gamma(\lambda)$ is a derived concept from the Jacobi equation (\ref{smile}) and this equation is well defined by the initial conditions at $\lambda=0$. The advantage of equation (\ref{smile2}) is that allows to state some continuity arguments which are useful for the present work. These arguments will be described in the next sections.

Another typical equation appearing in the literature \cite{marolf}
is given in terms of the expansion parameter $\theta_\gamma(\lambda)$, which is related to $G_\gamma(\lambda)$ by the formula
\be\lb{folio}
G_\gamma(\lambda)=G_i\exp \frac{1}{2}\int_{\lambda_i}^\lambda \theta_\gamma(\chi) d\chi,
\ee
with $G_i=G(\lambda_i)$ the value of $\sqrt{\det A_\gamma(\lambda)}$ at generic parameter value $\lambda_i>0$. 
In terms of $\theta_\gamma$ the equation (\ref{smile2}) becomes the well known Raychaudhuri equation
\be\lb{raychaudhuri}
\frac{d\theta_\gamma}{d\lambda}+\frac{\theta_\gamma^2}{2}=-\sigma_{\mu\nu}\sigma^{\mu\nu}-R_{\mu\nu}k^\mu k^\nu.
\ee
The definition (\ref{folio}) implies that
\be\lb{nana}
\theta_\gamma=\frac{2}{G_\gamma(\lambda)}\frac{dG_\gamma(\lambda)}{d\lambda}
\ee
Thus $\theta_\gamma(\lambda)\to-\infty$ when $\lambda\to \lambda_0$, since  $G_\gamma(\lambda)$ approaches to zero from positive values at  $q$.
Analogously, $\theta_\gamma(\lambda)\to\infty$ when $\lambda\to 0$, since  $G_\gamma(\lambda)$ grows from the zero value when starting at $p$.

On the other hand, the fact that $\theta_\gamma\to-\infty$ at $q$ itself does not imply that $G_\gamma(\lambda)\to 0$ when $\lambda\to\lambda_0$. This can be seen from (\ref{folio}), as the integral of 
the divergent quantity $\theta_\gamma$ may be still convergent. However, $G_\gamma(\lambda)$ should tend to zero when $\lambda\to\lambda_0$ by the very definition of conjugate point given below (\ref{smile}). By an elementary analysis of improper integrals it follows that, at the conjugate point $q=\gamma(\lambda_0)$, the expansion parameter $\theta_\gamma(\lambda)$ is divergent  
with degree
\be\lb{asin}
\theta_\gamma(\lambda)\sim \frac{-1}{|\lambda-\lambda_0|^{1+\epsilon}},\qquad \epsilon\geq 0,
\ee
up to multiplicative constant. The behavior (\ref{asin}) will play an important role in the next sections.

Note that the quantity (\ref{folio}) is not well defined when $\lambda_i \to 0$, that is, when the initial point is $p$. 
This reflects the expansion parameter is singular at $p$.

 Each of the equations (\ref{raychaudhuri}) and (\ref{smile2}) have their own advantages. In the following, both versions will play an important role, and will be employed in each situation by convenience.
\subsection{Future light cones in curved space times}
In addition to conjugate points, another important concept is the future light cone emanating from a point $p$ in the space time ($M$, $g_{\mu\nu}$). Given the point $p$ this cone is constructed in terms of all the future directed null vectors $k^\mu$ in $TM_p$ which satisfy the normalization (\ref{norma}).
Far away from $p$ these geodesics form a congruence  $\gamma_\sigma(\lambda)$, but for $\lambda=0$, the congruence is singular since $\gamma_\sigma(0)=p$ for every value of $\sigma$. In other words, $p$
is the tip of the cone.

Close to the point $p$ there is an open set $U$ composed by points $p'$, with their respective set of future directed null vectors $k'^\mu$  in $TM_{p'}$ which satisfy the normalization (\ref{norma}).
When comparing geodesics emanating from different points $p$ and $p'$, not only both points should be compared, but also the corresponding null vectors $k_\mu$ and $k'_\mu$. In some vague sense, two null geodesics $\gamma$
and $\gamma'$
are "close' when $p$ and $p'$ are close and the corresponding vectors $k_\mu$ and $k'_\mu$ "point in similar directions". 
In order to put this comparison in more formal terms, it is convenient to introduce the set $S$ defined as follows \cite{gaowald}
\be\lb{s}
S=\{\Lambda=(p, k^\mu)~|~ p \in M, \quad k^\mu \in TM_p,\quad k^\mu k_\mu=0, \quad k^\mu t_\mu=-1\}.
\ee
This set has an appropriate topology which allows to compare a pair $\Lambda=(p, k^\mu)$ with another one $\Lambda'=(p', k'^\mu)$ and to determine if they are ``close''. The definition implies that the vectors $k^\mu$ are all null and
satisfying the normalization (\ref{norma}).

\section{A continuity argument for GR with Null Conditions and its use}
\subsection{Statement and proof of the lemma}
The following lemma is of fundamental importance for the proof Gao-Wald statement in its original form \cite{gaowald}, and a similar lemma will be important for the present work. 
This lemma assumes that  the Null Energy and Null Generic Conditions are fulfilled and that the underlying theory is GR. As discussed in the introduction, these null conditions imply that every null geodesic in the space time contains at least a pair of conjugate points $p$ and $q$. 

%

Before describing the proof of this lemma, it is convenient to make a small redefinition of notation. In the following, the null geodesic defined by the pair $\Lambda=(p, k^\mu)$ will be denoted as $\gamma_\Lambda(\lambda)$. All the quantities depending on this curve such as $G_\gamma(\lambda)$
will be subsequently denoted as $G_\Lambda(\lambda)$ and so on. This notation is more adequate for studying the continuity properties of these quantities as functions on the space $S$ defined in (\ref{s})\footnote{Another convenient notation is to denote such quantities as $G(\lambda, \Lambda)$ and so on. We decided to use the notation described above instead.}.
\\

\emph{Lemma 1:} Assume that a given space time $(M, g_{\mu\nu})$  is described by the Einstein equations with matter content satisfying the Null Energy and Null Generic condition. Consider a pair $\Lambda_0=$($s_0$, $k_0^\mu$) in $S$,  and a pair of conjugate points $q_0$ to $p_0$ along $\gamma_{\Lambda_0}(\lambda)$. Then, there exists an open set $O$ in $S$ containing $\Lambda_0$ for which the following two properties hold.
\\

a) For every pair  $\Lambda=$($p$, $k^\mu$) in $O$, the corresponding geodesic $\gamma_\Lambda(\gamma)$ will posses at least a conjugate point $q_\Lambda$ to $p$,  $q_\Lambda \in J_+(p)-I_+(p)$. 
\\

b) The map $h: O\to M$ defined by $h(\Lambda)=q_\Lambda$, with $q_\Lambda$ the first conjugate point to $p$,
is continuous at $\Lambda_0$.
\\

The intuition of the lemma 1 is simply that, if one choses any point $p$ close enough to $p_0$ and draw a geodesic emanating from it with $k^\mu$ pointing
in a direction "similar" to $k^\mu_0$, then there will appear  a conjugate point $q$ to $p$ along this curve that is "very close" to $q_0$. That is roughly what the continuity statement is all about. The reason for it which becomes a bit technical is that the notion of being "very close" becomes tricky in non Riemannian geometry, as the distance between two widely separate points may vanish.
\\

\textit{Proof:} Denote the null geodesic $\gamma_{\Lambda_0}(\lambda)$ as $\gamma_0(\lambda)$ by simplicity, and choose the parameter $\lambda$ such that $p_0=\gamma_0(0)$ and $q_0=\gamma_0(\lambda_0)$ are conjugate points, with $\lambda_0>0$. This pair of conjugate points exists, as the Null Generic Condition is assumed to hold.
On the other hand, the results of the previous section show that $G_0(0)=G_0(\lambda_0)=0$
and $G_0(\lambda)>0$ for all $\lambda$ in the interval $0<\lambda<\lambda_0$. The Null Energy Condition $T_{\mu\nu}k^\mu k^\nu\geq 0$ implies, in the context of General Relativity, that $R_{\mu\nu}k^\mu k^\nu\geq 0$ for $k^\mu$ a null vector.
This, together with (\ref{smile2}) shows that $G_0''(\lambda)<0$ in the interval $0<\lambda<\lambda_0$.  The mean value theorem applied to $G_0$ shows that $G_0'(\lambda_1)=-C^2$ for some value $\lambda_1$
in the interval and furthermore  $G_0'(\lambda_1)<-C^2$ for $\lambda_1<\lambda<\lambda_0$, with $C^2$ a positive constant. By choosing $\lambda_0-\delta<\lambda<\lambda_0$ it is found that
\be\lb{hc1}
\frac{G_0(\lambda_1)}{|G'_0(\lambda_1)|}<\delta,
\ee
since $|G_0'(\lambda)|$ is larger than $C^2$ and $G_0(\lambda_1)$ is very close to zero if $\delta$ is small enough. 

Consider now a small open $O\subset S$ around the point $\Lambda_0=(p_0, k^\mu)$ generating $\gamma_0(\lambda)$.
As $G_\Lambda(\lambda)$ and its derivatives are continuous when moving in this open set, it follows that $G'_\Lambda(\lambda)<0$ and that
\be\lb{hc2}
\frac{G_\Lambda(\lambda_1)}{|G'_\Lambda(\lambda_1)|}<\delta,
\ee
for all the $\Lambda=(p, k^\mu)\in O$ if $O$ is small enough. What (\ref{hc1})-(\ref{hc2}) is showing is that the absolute value of the derivative
$G'_\Lambda(\lambda)$ is much more larger than $G_\Lambda(\lambda)$ in this small set. As the function $G_\Lambda(\lambda)$ has second derivative due to (\ref{smile2}), it follows that it is differentiable. This together with (\ref{hc2}) and the fact that $G_0'(\lambda_1)<-C^2$ and that $G''_\Lambda(\lambda)<0$ imply that
$$
G_\Lambda(\lambda_1)+G'_\Lambda(\lambda_1)\delta<0,
$$
This means that
$$
G_\Lambda(\lambda_1+\delta)+O(\delta)<0,
$$
with $O(\delta)$ going to zero faster than $\delta$. Thus $G_\Lambda(\lambda_2)=0$ for a $\lambda_2$ such that $|\lambda_2-\lambda_1|<\delta$.
This shows that there exists a conjugate point $q$ to $p$, which  is close to $q_0$ when $O$ is small enough. (Q.E.D)
\\

The lemma 1 is intuitive but technical in nature, for this reason it may be convenient at this point to explain its utility. This is done in the next subsection.
\subsection{The utility of the lemma 1}
Perhaps the best way to explain the use of the Lemma given above is to show how it leads to the Gao-Wald statement right from the scratch. 

As the manifold $M$ is by assumption paracompact, it can be made into a Riemannian manifold with Riemannian metric $q_{\mu\nu}$ \cite{Hicks}. 
The advantage of passing to this Riemannian setting is that
the the distance function between two points is small only if the points are close, while in a Minkowski space the condition of zero distance may not describe closeness adequately. 
The metric $q_{\mu\nu}$ can be converted into a complete one by a conformal transformation \cite{Hicks}, thus completeness may be assumed without further reasoning.  Fix any point $r \in M$ and let $d_{r}:M\to R$, with $d_{r}(s)$ the geodesic distance between $r$ and $s$ with respect to the metric $q_{\mu\nu}$. The function $d_r$ is continuous in $M$  and for all $\mathtt{R}>0$ the set $B_{R}=\{p\in M:~ d_{r}(p)\leq \mathtt{R}\}$ is compact (this follows from \cite[Theorem 15]{Hicks}).

Now, given $\Lambda \in S$, with $S$ the set defined in (\ref{s}) let $\gamma_{\Lambda}$ be the null geodesic determined by $\Lambda$. Consider the function $f:S\to R$ defined as 
$$f(\Lambda) =\{\inf_{R}B_R\;\; \textrm{such that contains a connected segment of}\;\; \gamma_{\Lambda}\;\; \textrm{that includes}
$$
$$
 \textrm{the initial point determined by }\;\Lambda\;\; \textrm{together with a pair of conjugate points of }\;\; \gamma_{\Lambda}\}.
 $$
In other words, $f(\Lambda)$ is constructed  starting with the point $\Lambda=$($s_0$, $k^\mu_0$) in $S$ by drawing the corresponding geodesic in the Riemannian geometry until two conjugated points have been found, and by finding a sort of minimal ball  $B_{R}$ that contains this drawing. Its radius $R$ is by definition $f(\Lambda)$. If there is no a line without conjugate points in the manifold, then it can be shown that $f(\Lambda_1)-f(\Lambda)<\epsilon$ for a $\Lambda$ in a open $O$ in $S$ containing $\Lambda_1$.  This is analogous to the condition of a continuity for a given function, but without the modulus. Such functions are called upper semicontinuous and have the property that they reach a maximum (but not necessarily a minimum) in a compact subset. The proof of the upper semi continuity property is given in the following lemma, see \cite{gaowald} and references therein for further details.
\\

\emph{Lemma 2:} The function $f(\Lambda)$ defined is upper semi continuous for a space time ($M$, $g_{\mu\nu}$) satisfying the Null Energy and Null Generic conditions. 
\\

\textit{Proof:}  Consider a geodesic $\gamma_0(\lambda)$ in the Minkowski geometry corresponding to a point $\Lambda_0=(s_0, k_0^{'\mu})$ in $S$. By hypothesis, this curve posses at least two conjugate points $p_0$ and  $q_0$. The parameter $\lambda$ may be chosen such that $p_0=\gamma_0(0)$. The tangent vector to the geodesic at this point satisfying (\ref{norma}) will be denoted as $k_0^\mu$. Analogously, let $\lambda_0$  the parameter corresponding to $q_0=\gamma_0(\lambda_0)$. The continuity lemma 1
implies that for any point $\Lambda=(p, k^\mu)$ in an open $O_1$ of $\Lambda_1=(p_0, k_0^\mu)$ small enough, there is a conjugate point $q=\gamma_\Lambda(\lambda_1)$ with $\lambda_0-\delta<\lambda_1<\lambda_0+\delta$. By making $O_1$ small enough, it may be shown that in the Riemannian geometry defined by the metric $q_{\mu\nu}$
the distance between $q_0$ and $q$ is less than $\epsilon_1$. 

Now, consider the parameter $\lambda'_0<0$ corresponding to $s_0=\gamma_0(\lambda'_0)$. Draw all the geodesics defined by the points $\Lambda$ in $O_1$ at the value $\lambda'_0$ together with their corresponding tangent null vectors $k^{'\mu}$. The points $s=\gamma_\Lambda(\lambda'_0)$ will lay inside a ball of radius $\epsilon_2$. By making $O_1$ small enough both $\epsilon_i<\epsilon/2$. By taking into account the continuity of the exponential map an open neighborhood $O_0$ of $\Lambda_0$ may be chosen, in such a way that the point $s$ will stay at distance less than $\epsilon/2$ of $s_0$ and the second conjugate points $q$ will be at a distance less than $\epsilon/2$ from $q_0$. This means that $f(\Lambda)<f(\Lambda_0)+\epsilon$ and this concludes the proof. (Q.E.D)
\\

\textit{Gao-Wald-Galloway proposition:} Let $(M,g_{\mu\nu})$ a space time satisfying the Null Energy and Null Generic conditions. Then, given a compact region $K$ in $M$ there exists a compact $K'$ containing $K$
such that, for any two points $p, q\notin K'$ and  $q$ belonging to $J_+(p)-I_+(p)$, no causal curve $\gamma$ joining $p$ with $q$ can intersect $K$.
\\

\emph{Proof:} The proof relies on the upper semi continuity property of lemma 2. Let $K\subset M$ be a compact set. Let $S_{K}=\{(p,k^{\mu})\in S,  p \in K\}$.  Since the tangent bundle has the product topology, $K$ is compact and $k^{\mu}$ is of bounded norm, it follows that $S_{K}$ is compact as well. Furthermore, as $f(\Lambda)$ is upper-semicontinuous, it must achieve a maximum $\overline{R}$ in $S_{K}$. Let  $K' = B_{\bar R}$ and let  $p,q\notin K'$ with $q\in J_+(p)-I_+(p)$. Construct a causal curve $\gamma$ joining $p$ with $q$. Then $\gamma$ must be a null geodesic since $q\in J_+(p)-I_+(p)$.  However, the Proposition 1 given in section 2 insures that  $\gamma$ should not contain a pair of conjugate points between $p$ and $q$. On the other hand, if $\gamma\cap K\neq \emptyset$ then by the definition of $K'$, $\gamma$ must have a pair of conjugate points lying in $K'$ and in between $p$ and $q$. This contradiction shows that $\gamma\cap K= \emptyset$ and therefore no of such causal curves cross $K$. This is the precisely the statement that was intended to be proved.(Q.E.D)
\\

It is important to remark that  all the proof given in the present section are based on the properties a) and b) of the Lemma 1, together with the absence of null curves that can not be deformed to time like curves. Thus, a generalization of these theorems for any matter field content and any  gravity theory is possible if these two properties of the lemma are satisfied, regardless the Null Energy and Null Generic conditions are relaxed or not.
 
\section{Generalization to general gravity models and arbitrary matter content}
The quantity $G_\Lambda(\lambda)$ defined in (\ref{marolf}) is crucial for the proof of Lemma 1 and for the proof of the Gao-Wald proposition. The fact that $G_\Lambda'(\lambda)$ is negative in certain region around the conjugate point $q_0$, and that  the inequality $G''_\Lambda(\lambda)<0$ is always satisfied, is of particular importance. This inequality is a consequence of the Null conditions, together with the Einstein equations. The aim of the present section is to generalize these statements to more general gravity models. The strategy is based on the observation that the desired generalization will hold if the conditions a) and b) of Lemma 1 are still satisfied. Therefore, it is important to understand if these two conditions are true for general models of gravity.

The problem with relaxing the Gao-Wald conditions and still be able to prove such continuity argument in terms of $G_\Lambda(\lambda)$
is the following. There is not obstruction to prove that, given a geodesic defined by some element $\Lambda_0$ in $S$, possessing  two conjugate points $p_0$ and $q_0$  (therefore $G_0(0)=G_0(\lambda_1)=0$), then for an open $O$ in $S$ small enough containing $\Lambda_0$ it follows that  $|G_\Lambda(\lambda_1)|<\delta$, with $\delta$ as small as possible.
However, this fact alone does not imply alone that there exists a pair of conjugate points $p=\gamma_\Lambda(0)$ and  $q=\gamma_\Lambda(\lambda_1+\epsilon)$ in $\gamma_\Lambda(\lambda)$, with $\epsilon$ small and $\Lambda$ a point in $O$. In fact, one may visualize $G_{\Lambda}(\lambda)$ as a sort of function in several variables, and there are a lot of such functions which has only isolated zeroes. If this is the case, the lemma 1 won't hold, as the property of possessing a pair of conjugate points will not be inherited by the "nearby" null geodesics of $\Lambda_0$. In other words, in order to generalize the Gao-Wald result, it must be show that the conjugate points do not "evaporate" by any slight perturbation of the curvature. This is not necessarily a trivial task.

Before going about this generalization, it should be mentioned that it seems possible for the authors to find a proof of this continuity statement for generic matter fields and generic gravity models, based on Morse theory \cite{Bott}-\cite{Milnor} and its applications to geodesics \cite{erlich}, \cite{Beem}-\cite{Uhlenbeck}. For achieving this proof,  a mathematical index form $I_\gamma$ associated to a given geodesic introduced in those references must be studied. By making a suitable Hilbert completion of the space of vector fields along a geodesic $\gamma$, the index form may be represented as an adjoint self operator in the corresponding Hilbert space. The problem of convergence of a set of geodesics could be then formulated as a convergence of the index of self-adjoint operators. This approach is not so direct however, as there exist sequences of symmetric matrices with limiting matrix with different index. This happens, for instance, if the limit matrix is not invertible. Thus, this idea has to be improved. One possibility is to consider an analogy between the problem of geodesics  with a Sturm Liouville problem of ordinary differential equations. The Sturm-Liouville  techniques are based on a bilinear form $B$ which can be related to the index of two homotopic curves in a real projective line. As the winding number is stable by homotopy, and in particular is protected by small perturbations, the index of  $B$ is stable by small perturbations. The references \cite{Helfer}-\cite{Mercuri} indicate that the correct index to be considered in the context of the Jacobi equation is the Maslov index. Given a particular geodesic $\gamma_0$, it is clear that a nearby geodesic $\gamma_\Lambda$ sees a slightly perturbed curvature term in the Jacobi equation. Then, based on the homotopy techniques described in those references, it may be possible that the number of conjugate points on $\gamma_\Lambda$ remain the same as in the unperturbed geodesic $\gamma_0$, and that they tend continuously to the ones of $\gamma_0$.

   One of the problems of the approach schematically described above is that is rather technical, although conceptually very rich and interesting. The other problem is that these references consider one parameter deformation of geodesics, while one may desire to consider a more general geodesic deformation. To generalize these theorems to this situation may require a lengthy proof, but we believe that this generalization is possible. Fortunately, we have found a way to overcome this difficulties and to proof the continuity lemma for general gravity and matter field contents. The proof is based on constructing a function $u_\Lambda(\lambda)$ analogous to the one $G(\Lambda, \lambda)$ described in previous section, for these general scenarios. The advantage is that this proof is readable for any researcher in gravity theory, without knowledge about the hard technical details described in the previous paragraph.   
   
The desired function may be constructed as follows. By redefining $\theta_\Lambda\to 2\theta_\Lambda$ for convenience, write the Raychaudhuri equation (\ref{raychaudhuri}) as
$$
\frac{d\theta_\Lambda}{d\lambda}=-\theta^2_\Lambda(\lambda)+p_\Lambda(\lambda), \qquad p_\Lambda(\lambda)=-2R_{\mu\nu}k^\mu k^\nu-2\sigma_{\mu\nu}\sigma^{\mu\nu}.
$$
By integrating this equation and taking the square of the result, one obtains that
$$
\theta_\Lambda^2=\bigg[-\int_{\lambda_0}^\lambda \theta^2_\Lambda(\xi)d\xi+p_\Lambda(\lambda)+\theta_{0\Lambda}\bigg]^2.
$$
By defining the quantity
\be\lb{posi}
R_\Lambda=\int_{\lambda_0}^\lambda \theta^2_\Lambda(\xi)d\xi, 
\ee
the last equation may be written in the following form
$$
\frac{dR_\Lambda}{d\lambda}=\bigg[-R_\Lambda(\lambda)+I_{\Lambda}(\lambda)\bigg]^2,\qquad I_\Lambda(\lambda)=p_\Lambda(\lambda)+\theta_{0\Lambda}.
$$
The definition of $R_\Lambda$ depends on an initial parameter $\lambda_0$. Note that the asymptotic behavior (\ref{asin})
implies that $R_\Lambda(\lambda)\to \infty$ when a conjugated point is reached.
By dividing by $R_\Lambda^2$ the following equation
\be\lb{poiseq}
\frac{d}{d\lambda}\bigg(\frac{1}{R_\Lambda}\bigg)=-\bigg[-1+\frac{I_{\Lambda}(\lambda)}{R_\Lambda}\bigg]^2,
\ee
is obtained. In terms of the quantity $u_\Lambda(\lambda)=R^{-1}_\Lambda(\lambda)$ the last equation becomes
$$
\frac{d u_\Lambda}{d\lambda}=-(u_\Lambda(\lambda)I_{\Lambda}(\lambda)-1)^2.
$$
The last formula implies that, regardless the gravity model or matter content in consideration, the derivative of $u_\Lambda=1/R_{\Lambda}$ is always less or equal to zero.
This observation will be of importance in the following.
 

The quantity $u_\Lambda(\lambda)=R^{-1}_{\Lambda}(\lambda)$ defined above is useful for proving the Lemma 1 for general space times. Note that the fact that $R_\Lambda\to \infty$ at the conjugate point implies that $u_\Lambda\to 0$ at this point.
\\

\emph{Lemma 3:} Given a space time $(M, g_{\mu\nu})$ consider a pair $\Lambda_0=$($s_0$, $k_0^\mu$) in $S$,  such that the corresponding geodesic  $\gamma_{\Lambda_0}(\lambda)$ contains a pair of conjugate points $p_0$ and $q_0$. Then, there exists an open set $O$ in $S$ containing $\Lambda_0$ for which the following two properties hold.
\\

a) For every pair  $\Lambda=$($p$, $k^\mu$) in $O$, the corresponding geodesic $\gamma_\Lambda(\gamma)$ will posses at least a conjugate point $q_\Lambda$ to $p$,  $q_\Lambda \in J_+(p)-I_+(p)$. 
\\

b) The map $h: O\to M$ defined by $h(\Lambda)=q_\Lambda$, with $q_\Lambda$ the first conjugate point to $p$,
is continuous at $\Lambda_0$.
\\

\emph{Proof:} Assume first that the function $I_\Lambda(\lambda)$ in (\ref{poiseq}) does not have any singularity.
By hypothesis, the null geodesic $\gamma_0(\lambda)$ contains two conjugate points $p_0=\gamma_0(0)$ and $q_0=\gamma_0(\lambda_1)$. 
Then it follows from (\ref{smile}) that there exist a matrix function $(A_\Lambda)_\mu^\nu$ satisfying 
$$
\frac{d^2A^\mu_\nu}{d\lambda^2}=-R^\mu_{\alpha\beta\gamma}k^\alpha k^\beta A^\gamma_\nu,
$$
together with the following initial conditions
$$
A^\mu_\nu|_{p_0}=0,\qquad \frac{dA^\mu_\nu}{d\lambda}\bigg|_{p_0}=\delta^\mu_\nu,
$$
such that $\det A|_{q_0}=0$. The initial data defines a point $\Lambda_0=(p_0, k^\mu_0)$ in $S$. For causal curves, there is an $\lambda_1>0$ such that $\gamma_0(\lambda)$ does not contains conjugated points to $p_0$ if $\lambda<\lambda_1$ \cite{oneil}.
That means that in this interval $\det A\neq 0$. Now, if the pair $\Lambda=(p, k^\mu)$ belongs to a small open $O$  containing $\Lambda_0=(p_0, k_0^\mu)$, then the same equation for $(A_\Lambda)^\mu_\nu$ may be solved along the geodesic $\gamma_\Lambda(\lambda)$, with exactly the same initial conditions. As the corresponding equation is linear and  $I_\Lambda(\lambda)$
is assumed to be non singular, it follows that if $O$ is small enough then the corresponding solution $(A_\Lambda)_\mu^\nu$ is continuous as a function of $\Lambda$. In other words, it is continuous as a function in compact set contained in $O$. Without losing generality the choice $\lambda_0=0$ may be employed for the initial points. As both  $\det A_\Lambda$ and its time derivative are continuous in $O$, the expansion parameter $\theta_\Lambda(\lambda)=(\det A_\Lambda)^{-1} \partial_\lambda \det A_\Lambda$ is continuous for $\lambda\neq 0$ as a function of $\Lambda$, if a conjugate point has not been reached.

 The continuity argument stated above implies that for  $\Lambda=(p, k^\mu)$ in open $O$ small  containing $\Lambda_0=(p_0, k_0^\mu)$ the inequality $|\theta_\Lambda(\lambda)-\theta_0(\lambda)|<\epsilon$ is valid, if $\lambda<\lambda_1$, with the size of $O$ depending on the choice of $\lambda$. On the other hand the definition (\ref{posi}) implies that
$$
R_\Lambda(\lambda)-R_0(\lambda)=\int_{\lambda_0}^\lambda [\theta^2_\Lambda(\xi)-\theta^2_0(\xi)]d\xi=\int_{\lambda_0}^\lambda [\theta_\Lambda(\xi)-\theta_0(\xi)][\theta_\Lambda(\xi)+\theta_0(\xi)]d\xi, 
$$
with $\lambda_0>\delta>0$. From here the following bound is found
$$
|R_{\Lambda}(\lambda)-R_0(\lambda)| \leq (\lambda-\lambda_0) Max (\theta_{\Lambda}+\theta_0)|_{[\lambda_0,\lambda]} \epsilon.
$$
As $\theta_\Lambda(\lambda)$ is finite in this interval, one may chose a compact $O'\in O$ and show that the last quantity is very small if $O$ and $O'$ are both small enough.
Thus $R_\Lambda(\lambda)$ is continuous in $O'$, unless a conjugated point is reached, since at this point the size of $O'$ becomes arbitrarily small. 

On the other hand, the function $u_\Lambda=1/R_\Lambda(\lambda)$ has in fact nicer properties than $R_\Lambda(\lambda)$. It is clearly continuous, even if $R_\Lambda$ blows up.
It follows then that if $u_0(\lambda_1)=0$, which implies that the conjugated point $q_0$ has been reached, then $u_\Lambda(\lambda_1)<\delta$ if $O$ in $S$ is small enough (note that $u_\Lambda>0$ since $R_\Lambda$ in (\ref{posi}) is obviously positive). As (\ref{poiseq}) shows that $u'_\Lambda(\lambda)\sim -1$ it follows that, for $\lambda<\lambda_1+\delta$,
$$
\frac{u_\Lambda(\lambda)}{|u'_\Lambda(\lambda)|}\leq \delta,
$$
if $O$ and $\delta$ are both very small. This inequality, and the differentiability of $u_\Lambda$ implies that $u_\Lambda(\lambda)=0$ for some $\lambda$ such that $|\lambda-\lambda_1|<\delta$. In other words, there is a conjugate point $q$
for any $\gamma_\Lambda$ originated by an initial data $\Lambda\in O$, for a parameter $\lambda_1-\epsilon<\lambda<\lambda_1+\epsilon$. This proves the theorem, if the curvature function $I(\lambda)$ does not achieve a singularity inside this set.

The other possibility is that $I_\Lambda(\lambda)$ is divergent at some point $\lambda_2$ of the curve $\gamma_0(\lambda)$. Then (\ref{poiseq}) shows that $u'_\Lambda \to -\infty$ when $\lambda\to \lambda_2$.
After some reasoning, it may be concluded that $u_\Lambda(\lambda_1)=0$ for $\lambda_1<\lambda_2$ and thus there will appear a conjugate point $q_0=\gamma_0(\lambda_1)$ before reaching the singularity. The discussion of the previous paragraph shows that the continuity argument still holds. This concludes the proof. (Q.E.D)
\\

 By virtue of lemma 3 it follows that, once a null geodesic contains two conjugate points $q_0$ and $p_0$, then the "nearby" null geodesics will also contain such pair $p$ and $q$. One can move along these null geodesics and  conclude that all of them have conjugate points unless the conjugate point $q$ moves far away. Another possible situation is that the point $q$ moves to a point where $I_{\Lambda}(\lambda)$ is finite but the space time is not extended further. Note that there exist space times which does not extend beyond some point or hypersurface even if there are no curvature singularities there \cite{geroch}. If anything like this happens, it  should not be necessarily concluded that the geodesic does not contains conjugate points. It may be the case that this null geodesic admits another pair of conjugate points $r_0$ and $s_0$ which are simply not "close" to $p_0$ and $q_0$. The nearby geodesics however, will also contain close conjugate pairs $r$ and $s$. One may wander again along the null geodesics around this particular one. By repeating this procedure, one may find eventually an inextensible null geodesic $\gamma:[a, b]\to M$ that has no conjugate points. If this geodesic appears, proposition 1 shows that it can not be deformed to a time like one. Its image is achronal. If instead this null line is not present in the space time, then the points a) and b) of Lemma 1 will be satisfied and the function $f(\Lambda)$ of lemma 2 becomes upper semi-continuous. Then the generalization of the Gao-Wald proposition follows, as it is a consequence of this upper semi continuity property and the presence of conjugate points in every null geodesics. In other words, the following has been proved.
 \\
 
\emph{Galloway statement:} Let $(M,g_{\mu\nu})$ be a generic space time. Then one of the two possibilities at least is realized.
\\

a) There is an inextensible null geodesic $\gamma:[a,b]\to M$ with achronal image (a null line).
\\

b) Given a compact region $K$ in $M$ there exists a compact $K'$ containing $K$
such that, for any two points $p, q\notin K'$ and  $q$ belonging to $J_+(p)-I_+(p)$, no causal curve $\gamma$ joining $p$ with $q$ can intersect $K$.
\\

Note that the possibility a) does not exclude the possibility b) or viceversa. In other words a) and b) can be realized simultaneously. The only thing that is not possible is to exclude both possibilities, at least one should be realized.

\section{Applications}
In the present work an original proof of the continuity Lemma  3, valid regardless the underlying theory or matter content, was presented. This lemma is not trivial from the mathematical or physical point of view, and is fundamental for extending the Gao-Wald statement about apparently faster than light travels to general gravity models with general matter content. In particular, it was shown that this statement leads directly to the Galloway unpublished statement. The presented arguments avoid the use of Morse theory, and is more accesible to researchers in gravity models.
The idea of the proof is to engine a suitable function $u_\Lambda(\lambda)$ analogous the Flanagan-Marolf-Wald function $G(\Lambda, \lambda)$, adapted to the more general context. 

The next task is to discuss some applications of the presented results. The results to be discussed below follow from \cite{gaowald} combined with the ones presented here. In that reference, several results 
follow directly from the Lemma 3. However, these authors present a proof of this lemma that is valid only for GR with Null Energy and Null Generic conditions. As this lemma was generalized here
to the more general context, analogous affirmations may be found for general gravity scenarios. As the discussion of all the underlying
physics would be lengthy, the exposition will be very succint. For details about the physical meaning of these statements we refer the reader to \cite{gaowald} and references therein.

The first application is the following proposition about the absence of particle horizons, if a null line is absent. This is the statement of Corollary 1 of \cite{gaowald} but with the  mentioned conditions replaced by the absence of a null line.
\\

\emph{Theorem 1:}
Let $(M, g_{\mu\nu})$ an space time without inextendible null geodesics with achronal images. If $(M, g_{\mu\nu})$ is globally hyperbolic with a compact
Cauchy surface $\Sigma$, then there exist Cauchy surfaces $\Sigma_1$
and $\Sigma_2$ (with $\Sigma_2\subset I^+(\Sigma_1$)) such that if
$q \in I^+(\Sigma_2)$, then $\Sigma_1 \subset I^-(q)$.
\\

Another important result is related to conformal embeddings of space times. Suppose that $(M,g_{\mu\nu})$ can be conformally embedded into another 
space time $(\widetilde M, \tilde g_{\mu\nu})$, so that in $M$ the relation
$\tilde{g}_{\mu\nu}=\Omega^2 g_{\mu\nu}$ holds. The boundary $\dot{M}$ of $M$ in $\widetilde M$ is assumed to be a time like hypersurface.  Given a point $p$ in $\dot{M}$, a set of interest is is composed by the points of $\dot{M}$ that can be joined by curves starting from $p$ and lying inside $M$, except for the endpoints. This motivates the following definition
\begin{eqnarray}
A(p)&=&\{ r \in \dot{M}| \mbox{there exists a future directed causal curve $\lambda$
starting} \nonumber \\
&& \mbox{from $p$ and ending at $r$ satisfying 
$\lambda - p\cup r \subset M\}$}.
\label{Ap}
\end{eqnarray}
The boundary of $A(p)$ in $\dot M$ is denoted as $\dot A(p)$. In these terms, by replacing the Gao-Wald conditions by the absence of null lines in Theorem 2 of reference \cite{gaowald}, the following statement is found.
\\

\emph{Theorem 2:}
Consider a space time $(M,g_{\mu\nu})$ that can be conformally embedded into another 
$(\widetilde M, \tilde g_{\mu\nu})$, so that in $M$ the relation
$\tilde{g}_{\mu\nu}=\Omega^2 g_{\mu\nu}$ holds, and on $\dot{M}$ the conformal factor $\Omega =
0$, where $\Omega$ is a smooth function on $\widetilde{M}$.  Assume that $(M,g_{\mu\nu})$
satisfies the following conditions.
\\

a)  $(M,g_{\mu\nu})$ does not contain a null line.
\\

b) $\bar M$ is strongly causal.
\\

c) For any $p, q\in \overline{M}=M\cup \dot{M}$, $J^+(p) \cap J^-(q)$ is a compact set. 
\\

d) $\dot M$ is a timelike hypersurface in $\widetilde{M}$. 
\\

Given a point $p\in \dot M$, for any $q\in \dot{A}(p)$, it follows that $q\in J^+(p)-I^+(p)$. Furthermore, any causal curve in $\bar{M}$
connecting $p$ to $q$ must lie entirely in $\dot M$ and, hence, must
be a null geodesic in the spacetime $(\dot M, \tilde{g}_{\mu\nu})$.
\\

In the statement of this theorem, all the past and future sets are taken with respect to $\overline{M}$.

It should be mentioned that Anti de Sitter spaces do not satisfy the hypothesis of this result.
However, any deformation of them which fails to produce a null line and satisfy all these hypothesis always produce time delay with respect to anti de Sitter
itself. This follows from the fact anti de Sitter space times admits pairs of point $p$ and $q$ connected by null geodesics which lies in $M$, but any of such deformation will move these geodesics to the boundary $\dot{M}$. The boundary is not changed for this geometry by the conformal transformation. Thus, the race between two null geodesics joining $p$ and $q$ is favored by the geodesic of the boundary, once the deformation takes place. It may be said that for these deformations of anti de Sitter space, there is a time delay with respect of the anti de Sitter space itself \cite{gaowald}. 

The results presented here are technical details that play an important role for the Penrose-Sorkin-Woolgar positive mass theorem \cite{sorokin}.
The authors \cite{sorokin} in fact realize that this theorem can be proved for more general theories than GR if the underlying space time if certain focusing conditions for null geodesics such as  \cite{tippler}-\cite{borde} are satisfied. These focusing conditions are chosen in order to assure that null lines are absent. More general conditions have been found in \cite{galloway1}-\cite{galloway2}. These conditions involve non local quantities constructed in terms of the curvature such as 
$$
I_\Lambda(\lambda_i)=\lim_{\lambda\to\infty} \textrm{inf}\int_{\lambda_i}^\lambda e^{-c \xi} [R_{\mu\nu}k^\mu k^\nu+\sigma_{\mu\nu}\sigma^{\mu\nu}]_\Lambda(\xi)d\xi,
$$
with $c_\Lambda>0$. It can be shown that if these quantities are not divergent, which in particular implies that a light traveller never finds an asymptotic exponential grow of the form $R_{\mu\nu}k^\mu k^\nu\sim -e^{c\xi}$, then given some suitable  initial conditions the presence of conjugate points may be insured, regardless the underlying gravity theory. We refer the reader to the original references, and to \cite{nova1}-\cite{nova3} for more information about this focusing conditions and for further applications.

As a final comment, we would like to mention that the results of the present work may be applied to the causality issues deeply studied in references \cite{caus1}-\cite{caus12}.
But these applications will be considered in a separate work.

\section*{Acknowledgments}
Both authors are supported by CONICET, Argentina.


\begin{thebibliography}{99}

\bibitem{alcubierre}  M. Alcubierre, Class. Quantum Grav. 11 (1994) L73.


\bibitem{gaowald} S. Gao and R. Wald Class. Quant. Grav 17 (2000) 4999.
\bibitem{krasnikov}  S. V. Krasnikov, Phys. Rev. D 57 (1998) 4760.

\bibitem{olum} K. Olum, Phys. Rev. Lett. 81 (1998) 3567.
\bibitem{otras} M. Visser, B. Bassett, and S. Liberati, Nucl. Phys. B88 (Proc. Supl) (2000) 267.
\bibitem{otras1} M. Visser, B. Bassett, and S.
Liberati, in General Relativity and Relativistic Astrophysics, Proceedings of the Eighth Canadian Conference, ed. by C.P Burgess and R.C. Meyers, (AIP Press, Melville, New York, 1999).
\bibitem{otras2} X. Camanho, J. Edelstein, J. Maldacena and A. Zhiboedov JHEP 02 (2016) 20.
\bibitem{otras3} G. Papallo and H. Reall JHEP 11 (2015) 109.
\bibitem{otras0}  X. Camanho, J. Edelstein and A. Zhiboedov Int. J. Mod. Phys. D 24 (2015) 1544031.
\bibitem{otras4}  J. Edelstein, G. Giribet, C. Gomez, E. Kilicarslan, M. Leoni and B. Tekin Phys. Rev. D 95 (2017) 104016.


\bibitem{tipler1} F. Tipler, Phys. Rev. Lett. 37 (1976) 879.
\bibitem{tipler2} F. Tipler, Ann. Phys. 108  (1977) 1.
\bibitem{tipler3} S. W. Hawking, Phys. Rev. D 46 (1992) 603.

\bibitem{Wald} R.M. Wald, General Relativity, University of Chicago Press (Chicago, 1984).
\bibitem{hawking} S.W. Hawking and G.F.R. Ellis, The Large Scale Structure of SpaceTime, Cambridge University Press (Cambridge, 1973).
\bibitem{erlich} J. Beem, P. Ehrlich and K. Easley Global Lorentzian Geometry CRC press 1981.
\bibitem{oneil} B. O Neill Semi-Riemannian Geometry with Applications to General Relativity Academic Press
1983.
\bibitem{penrose} R. Penrose, in Essays in General Relativity, ed. by F.J. Tipler, Academic
Press (New York, 1980).
\bibitem{odintsov} S. Nojiri, S. Odintsov and V. Oikonomou Phys. Rept. 692 (2017) 1.
\bibitem{odintsov1} K. S. Stelle, Phys. Rev. D 16 (1977) 953.
\bibitem{odintsov2} K. S. Stelle, Gen. Rel. Grav. 9 (1978) 353.
\bibitem{odintsov3}  D. Lovelock  J. Math. Phys. 12 (1971) 498.


\bibitem{averaged1} H. Epstein, V. Glaser, and A. Jaffe Nuovo Cim 36 (1965) 1016.
\bibitem{averaged2} C. Fewster Class. Quant. Grav. 17 (2000) 1897.
\bibitem{averaged1} C.  Fewster and S. Eveson  Phys. Rev. D 58 (1998) 084010.
\bibitem{averaged3}  C. Fewster and S. Hollands Rev. Math. Phys. 17 (2005) 577.
\bibitem{averaged4} C. Fewster, K. Olum, and M. Pfenning Phys. Rev. D 75 (2007) 025007.
\bibitem{averaged5} C. Fewster and L. Osterbrink  Phys. Rev. D 74 (2006) 044021.
\bibitem{averaged6} C. Fewster and L. Osterbrink J. Phys. A 41 (2008) 025402.
\bibitem{averaged7} C. Fewster and T. Roman Phys. Rev. D 67 (2003) 044003.
\bibitem{averaged8} C. Fewster and C. Smith Annales Henri Poincare 9 (2008) 425.
\bibitem{averaged9} L. Ford and T. Roman
Phys. Rev.  D 51 (1995) 4277.
\bibitem{averaged10} L. Ford and T. Roman Phys. Rev. D 53 (1996) 5496.
\bibitem{averaged11} N. Graham and K. Olum Phys. Rev. D 76 (2007) 064001.
\bibitem{averaged13} E. Kontou and K. Olum Phys. Rev. D 87 (2013) 064009.
\bibitem{averaged15} E. Kontou and K. Olum Phys. Rev. D 90 (2014) 024031.


\bibitem{averaged16} D. Urban and K. Olum Phys. Rev. D 81 (2010) 024039.
\bibitem{averaged17} D. Urban and K. Olum Phys. Rev. D 81 (2010) 124004.

\bibitem{averaged19} M. Visser Phys. Lett. B 349 (1995) 443.
\bibitem{averaged20} M. Visser Phys. Rev. D 54 (1996) 5103.
\bibitem{averaged21} M. Visser Phys.Rev. D 54 (1996) 5116.
\bibitem{averaged22} M. Visser  Phys. Rev. D 56 (1997) 936.
\bibitem{averaged23} R. Wald Phys. Rev. D 17 (1978) 1477.
\bibitem{averaged24} F. J. Tipler Phys. Rev. D17 (1978) 2521.
\bibitem{averaged25} A. Borde Class. Quant. Grav. 4 (1987)
343.
\bibitem{averaged26} R. Bousso, Z. Fisher, S. Leichenauer, and A. C. Wall Phys. Rev.
D93 (2016) 064044.
\bibitem{averaged28} S. Balakrishnan, T. Faulkner, Z. U. Khandker, and H. Wang  JHEP 09 (2019) 020.
\bibitem{averaged29} J. Koeller and S. Leichenauer Phys.
Rev. D 94 (2016)  024026.
\bibitem{averaged30} F. Rosso JHEP 03 (2020) 186.
\bibitem{marolf} E. Flanagan, D. Marolf, and R. Wald Phys. Rev. D 62 (2000) 084035.

\bibitem{average1} A. Borde, Class. Quant. Grav. 4 (1987) 343.
\bibitem{galloway1}  C. Fewster and G. Galloway Class. Quantum Grav. 28 (2011) 125009.
\bibitem{galloway11} P. Brown, C. Fewster and E. Kontou Gen Relativ Gravit. 50 (2018) 121.
\bibitem{galloway2} G. Galloway Math. Proc. Camb. Phil. Soc (1986) 99367.
\bibitem{Hicks} N. Hicks, Notes on Differential Geometry, Van Nostrand (Princeton 1965).
\bibitem{geroch} R. Geroch Annals of Physics 48 (1968) 526.
\bibitem{sorokin} R. Penrose, R. D. Sorkin, and E. Woolgar, A positive mass theorem based on the focusing and retardation of null geodesics, preprint gr-qc/9301015, 1993; R. Penrose  Twistor Newsletter 30 (1990), 1; R. D. Sorkin and E. Woolgar, New demonstration of the positivity of mass, in: Proc. Fourth Can. Conf. on Gen. Rel. and Rel. Astrophys. , G. Kunstatter, D. E. Vincent, and J. G. Williams (eds.), World Scientific, Singapore, 1992, 206. 
\bibitem{borde} A. Borde Class. Quantum Gravit. 4 (1987) 343.
\bibitem{tippler} F. Tipler Phys Rev D 17 (1978) 2521.
\bibitem{nova1} G. Galloway, S. Suruya and E. Woolgar Commun.Math.Phys. 241 (2003) 1.
\bibitem{nova2} G. Galloway Lect.Notes Phys. 604 (2002) 51.
\bibitem{nova3} G. Galloway and D. Solis Class.Quant.Grav. 24 (2007) 3125
\bibitem{null} G, Galloway Annales Poincare Phys.Theor. 1 (2000) 543.
\bibitem{Bott} R. Bott  Commun. Pure Appl. Math. 9 (1956), 171–206.
\bibitem{Smale} S. Smale, On the Morse Index Theorem, J. Math. Mech. 14 (1965), 1049–1056.
\bibitem{Milnor} J. Milnor, Morse Theory, Princeton Univ. Press, Princeton, 1969.
\bibitem{Beem} J. Beem, P. Ehrlich Duke Math. J. 46 (1979), 561– 569.
\bibitem{Woodhouse} N. Woodhouse Comm. Math. Phys. 46 (1976) 135.
\bibitem{Uhlenbeck} K. Uhlenbeck Topology 14 (1975) 69.

\bibitem{Helfer} A. Helfer Pacific J. Math. 164, n. 2 (1994), 321.
\bibitem{Helfer2} A. Helfer Contemporary Mathematics 170 (1994), 135..
\bibitem{Mercuri} F. Mercuri, P. Piccione, and D. Tausk Pacific Journal of Mathematics, vol. 206, No. 2 (2002)  375.
\bibitem{caus1} G. M. Shore, Int. J. Mod. Phys. A 18 (2003) 4169.
\bibitem{caus2} G. M. Shore, Nucl. Phys. B 778 (2007) 219.
\bibitem{caus3} T. J. Hollowood and G. M. Shore, Phys.
Lett. B 655 (2007) 67.
\bibitem{caus11} T. Hollowood and G. Shore J. Phys. A: Math. Theor. 49 (2016) 215401.
\bibitem{caus4} T. J. Hollowood and G. M. Shore, Nucl. Phys. B 795 (2008) 138.
\bibitem{caus5} T. J. Hollowood and G. M. Shore, JHEP 0812 (2008) 091.
\bibitem{caus6} T. J. Hollowood, G. M. Shore and R. J. Stanley, JHEP 0908 (2009) 089.
\bibitem{caus7}  T. J. Hollowood and G. M. Shore, Phys. Lett. B 691 (2010) 279.
\bibitem{caus8} T. J. Hollowood and G. M. Shore,  JHEP 1202 (2012) 120.
\bibitem{caus9} C. de Rham and A. J. Tolley, Phys. Rev. D 101 (2020) 063518.
\bibitem{caus10} C. de Rham and A. J. Tolley,
Phys. Rev. D 102 (2020) 084048.
\bibitem{caus12} C. de Rham, S. Melville and J. Noller JCAP 08 (2021) 018.
\end{thebibliography}
\end{document}